\documentclass[conference]{IEEEtran}
\IEEEoverridecommandlockouts
\usepackage{cite}
\usepackage{amsmath,amssymb,amsfonts}
\usepackage{algorithmic}
\usepackage{graphicx}
\usepackage{textcomp}
\usepackage{xcolor}
\usepackage{listings}
\newcommand{\ie}{{\it i.e.,}}

\begin{document}

\title{An empirical study on Java method name suggestion: are we there yet?\\
\thanks{}
}

\author{\IEEEauthorblockN{Weidong Wang}
\IEEEauthorblockA{\textit{Faculty of Information Technology} \\
\textit{Beijing University of Technology}\\
Beijing, China \\
wangweidong@bjut.edu.cn}
\and
\IEEEauthorblockN{Yujian Kang}
\IEEEauthorblockA{\textit{Faculty of Information Technology} \\
\textit{Beijing University of Technology}\\
Beijing, China \\
kangyujian@gmail.com}
\and
\IEEEauthorblockN{Dian Li}
\IEEEauthorblockA{\textit{Faculty of Information Technology} \\
\textit{Beijing University of Technology}\\
Beijing, China \\
iamlidian@emails.bjut.edu.cn}
}

\maketitle

\begin{abstract}
	
A large-scale evaluation for current naming approaches substantiates that such approaches are accurate. However, it is less known about which categories of method names work well via such naming approaches and how's the performance of naming approaches. To point out the superiority of the current naming approach, in this paper, we conduct an empirical study on such approaches in a new dataset. Moreover, we analyze the successful naming approaches above and find that: 1) around 60\% of the accepted recommendation names are made on prefixes within \emph{get}, \emph{set}, \emph{is}, and \emph{test}. 2) a large portion (19.3\%) of method names successfully recommended could be derived from the given method bodies. The comparisons also demonstrate the superior performance of the empirical study.

\end{abstract}


\section{Introduction}
\label{sec1}

In software project, while coding a function \/or method body, developers give a name to the body as method name. A good method name not only helps programmers understand the program code better\cite{g28,g29}, but also prevents method APIs from being operated incorrectly \cite{g27}. Since function call using specified method name is commonplace in program code, more and more software companies emphasize method naming conventions and programming standards. As emphasized by the industrial experts including McConnell \cite{g6} and Beck \cite{g6-1}, method naming is one of the significant software activities.

While there are an army of implied standards for method naming such as the CamelCase principle \cite{g6-2,g30}, there is no broad understanding of which categories of method names are prone to be used or accepted by developers. We investigate the current naming approaches \cite{g11,g3,g1,g13} in the past three years and find that in the correctly suggested method names, the categories within prefixes such as \emph{get}, \emph{set}, \emph{is}, and \emph{test} account for 60\%. If we remove such method names within the prefixes above, the precision and recall of suggested method names could be downgraded dramatically. Actually, in coding process, we commonly need to name a potential method body that is far beyond the category including the above prefixes.

Furthermore, we don't know which naming approach is outstanding under the same experiment environment since the evaluation metrics of such naming approaches are not uniform as well \cite{g31}. For example, the actual method name suggested is likely to be locally different from the given method name, however, the meanings of both method names are the same. To evaluate the above method names suggested, \emph{code2seq} \cite{g11} and \emph{code2vec} \cite{g13} redefined the loose evaluation metrics, while the \emph{HeMa} \cite{g1} adopts tightened evaluation metrics \ie\ F1-score, recall, and precision. Meanwhile, the \emph{HeMa} approach also uses panel review to evaluate the method names suggested by inviting domain experts.

Although such approaches are overall accurate in suggesting method name, little is known about which categories and why they work or don't work. And we also know little about the actual performance on a new dataset. Indeed, it is hard to automatically generate high-quality method name including answering why an accurate method name occurring frequently requires synthesis of different kinds of knowledge and context. Hence, we conduct emprical study among those famous approaches related with method name suggestion. 

The rest of this paper is organized as follows. Section \ref{sec2} introduces the new datasets. Section \ref{sec3} describes the experiment results. Section \ref{sec4} investigates the related work. Section \ref{sec5} concludes this paper.

\section{New Datasets}
\label{sec2}

To make the following empirical study more objective and fair, we built 3 datasets for the purpose of experimental evaluation namely ``Liu-kui dataset", ``new-100 dataset", and ``new-400 dataset", respectively. The details of these datasets are as follows.

\begin{itemize}
	\item Liu-kui dataset: This dataset was collected from 4 highly rated open source organizations, Apache, Spring, Hibernate and Google. 430 projects with the latest versions of Java source code were employed, each of which has 100 times submissions at least. This dataset totally contains 2,116,413 methods. The further details of the dataset can be found in Liu's literature \cite{g2}.
	\item New-100 dataset: We randomly selected 100 projects from the open source Java projects on the former 10K GitHub, and each of these projects has been submitted at least 100 times. The dataset contains 47,098 Java methods.
	\item New-400 dataset: We randomly selected nearly 4.8 million methods from the open source Java projects on the former 10K GitHub to form this dataset.
\end{itemize}

The above three datasets have been preprocessed by filtering \emph{main method}, \emph{constructor}, \emph{easy method}, \emph{empty method}, and \emph{method with no letters} in method names since these methods do not often cause trouble for the maintenance and understanding of programs, but interfere with the naming of methods.

\section{Empirical Study}
\label{sec3}

To evaluate the performance of such naming approaches, we conducted a series of experiments. The following parts mainly consist of the parameter settings of experiments, the choices of evaluation metrics, and the answers to the research questions from RQ1 to RQ3. 

1) In order to study the advanced approaches to Java method name recommendation in this paper, we propose the following research questions (RQs) as follows.

\begin{itemize}
  \item \textbf{RQ1}: Can \emph{code2seq} perform as well on the new dataset as the original dataset in this paper?
  \item \textbf{RQ2}: If the AST paths of a method do not contain a token in the method name, what is the performance?
  \item \textbf{RQ3}: Can the length of the method names recommended fulfill the standard of method name length?
\end{itemize}

To ensure the accuracy of the experimental results, we divide the training set and the test set based on 10-cross validation. In the evaluation process, we take the average of the results with 10-cross validation as the final result.

\begin{table}[h]
\center
\begin{tabular}{|c|c|c|}
\hline
\multicolumn{3}{|c|}{[Dataset 1] Liu-kui Dataset} \\ \hline
Test          & Training          & Total            \\ \hline
211641        & 1904772        & 2,116,413        \\ \hline
\multicolumn{3}{|c|}{[Dataset 2] New-100 Dataset} \\ \hline
Test          & Training          & Total            \\ \hline
4710          & 42388          & 47098            \\ \hline
\multicolumn{3}{|c|}{[Dataset 3] New-400 Dataset} \\ \hline
Test          & Training         & Total            \\ \hline
481039        & 4329351        & 4810390          \\ \hline
\end{tabular}
\caption{The details of training and test samples in the three datasets.}
\label{tab-1}
\end{table}

\noindent \textbf{RQ1: Can \emph{code2seq} perform as well on the new dataset as the original dataset in this paper?}

The \emph{code2seq} was initially evaluated on the original dataset. The experimental results show that the precision, recall, and F1-score reach 64.3\%, 55.02\%, 67.1\% \cite{g11}. The \emph{code2seq} is evaluated on the three datasets above, namely Liu-kui dataset, new-100 dataset, and new-400 dataset. The evaluation metrics utilized by \emph{code2vec} are employed as the unified evaluation metrics in this paper due to its efficiency.

For a given pair of real method name $e$ and recommended method names $r$, we define the precision as $pre(e,r)$, recall as $rec(e,r)$, $pre(e,r)=\frac{|token(e)\bigcap token(r)|}{token(r)}$, and $rec=\frac{|token(e)\bigcap token(r)|}{token(e)}$. $token(n)$ returns the tokens in the method name $n$.

The evaluation results are in the following table. As in \emph{code2vec}, we choose Precision, Recall and F1-score as evaluation indexes. The right three columns correspond to the performance of \emph{code2seq} on the three different datasets. Precision, Recall, and F1-score in the table are the average values of each item in the test set.

\begin{table}[h]
\center
\begin{tabular}{|c|c|c|c|}
\hline
          & \textbf{new-400 dataset}  &\textbf{ Liu-kui dataset } & \textbf{new-100 dataset} \\ \hline
\textbf{Precision} & 0.70             & 0.67             & 0.64            \\ \hline
\textbf{Recall}    & 0.63             & 0.56             & 0.53            \\ \hline
\textbf{F1-score } & 0.66             & 0.60              & 0.58            \\ \hline
\end{tabular}
\caption{The performance of \emph{code2seq} on the three datasets.}
\label{tab:my-table}
\end{table}

From the table, we have the following findings:

\begin{itemize}
  \item The volume of datasets may have an influence on the performance of \emph{code2seq}. Changing the volume of new-400 dataset, Liu-kui dataset, new-100 dataset, we can find that the performance of \emph{code2seq} improves significantly with the increase of dataset volume.
  \item Compared with Liu's approach \cite{g2}, the precision, recall and F1-score of \emph{code2seq} on the largest new-400 dataset are improved by 3\%, 7\% and 6\% respectively, while the precision, recall and F1-score of \emph{code2seq} on the new-100 dataset are improved by 6\%, 10\%, and 8\%, respectively.
\end{itemize}

From the above experiments and analysis, we can draw the conclusion that \emph{code2seq} will still keep certain accuracy after transforming datasets. Nevertheless, different datasets have great influence on the performance of \emph{code2seq}, and its performance on larger dataset is obviously better than that on smaller datasets due to the significant increase of training samples on larger datasets.

\noindent \textbf{RQ2: If the AST paths of a method do not contain a token in the method name, what is the performance?}

To study RQ2, we evaluated the performance of \emph{code2seq} on visible and invisible methods. We will give the definitions of visible method and invisible method as follows.

For each method name $n_i$, in the Liu-kui dataset, we divide it into a token stream according to the CamelCase principle, $T_i=\{t_1,t_2,t_3,\cdots,t_m\}$,

where $t_i$ is the $i$-th token in the token stream. The method body $b_i$ corresponding to the method name $n_i$ is parsed into an army of AST paths, where $p_j$ is the $j$-th AST path, consisting of several tokens, $p_j=\{s_1,s_2,\cdots,s_k\}$, and $s_k$ is the $k$-th token. $Q_i$ is the set of all AST path tokens belonging to $b_i$, $Q_i=\{p_1,p_2,\cdots,p_j\}$. If each token in $T_i$ does not appear in $Q_i$, $n_i$ is an invisible method, otherwise, it's a visible method.

Next, we evaluate the performance of \emph{code2seq} on both visible and invisible methods, we select 918k invisible methods from the Liu-kui dataset as a new dataset called ``invisible dataset''. Accordingly, we randomly choose the same number of visible methods from the Liu-kui dataset to form the corresponding ``visible dataset''. On both datasets above, we evaluate the \emph{code2seq}'s performance via precision, recall and F1-score metrics.

\begin{table}[h]
\center
\begin{tabular}{|c|c|c|}
\hline
            & \textbf{Invisible dataset} & \textbf{Visible dataset} \\ \hline
\textbf{Total}   & 918545            & 918545       \\ \hline
\textbf{Correct} & 250763            & 387626       \\ \hline
\textbf{Correct rate}      & 0.273             & 0.422        \\ \hline
\end{tabular}
\caption{The correct rate on the visible and invisible datasets.}
\label{tab:my-table}
\end{table}

Here, we have several findings as follows.

\begin{itemize}
  \item First of all, as shown in Table \ref{tab:1-2}, a large proportion of method name tokens (accounting for 43\%) can not be found in their corresponding method bodies. Therefore, \emph{code2seq} based on machine learning is often limited by its replication mechanism and vocabulary size, which results in performance decline.

\begin{table}[h]
\center
\begin{tabular}{|c|c|c|}
\hline
       & \textbf{Total}    & \textbf{Invisible dataset} \\ \hline
\textbf{Number} & 2116413 & 918545         \\ \hline
\textbf{Rate}   & \multicolumn{2}{c|}{0.43} \\ \hline
\end{tabular}
\caption{The proportion of the number of invisible methods to the total number of methods.}
\label{tab:1-2}
\end{table}

  \item What's more, even under the limitation mentioned in the first point, \emph{code2seq} still performs well in the field of method name recommendation on the invisible datasets. As shown in Table \ref{tab:1-3}, it maintains 39\% precision, 37\% recall and 38\% F1-score. Compared with \emph{code2seq}'s recommendation effect on the visible methods, the corresponding precision, recall, and F1-score only have a significant decrease of 35\%, 25\%, and 29\%, respectively.

\begin{table}[b]
\center
\begin{tabular}{|c|c|c|}
\hline
          & \textbf{Invisible dataset} & \textbf{Visible dataset} \\ \hline
\textbf{Precision} & 0.39                    & 0.74                  \\ \hline
\textbf{Recall}    & 0.37                    & 0.62                  \\ \hline
\textbf{F1-score}  & 0.38                    & 0.67                  \\ \hline
\end{tabular}
\caption{The performance of \emph{code2seq} on the visible and invisible datasets.}
\label{tab:1-3}
\end{table}
\end{itemize}

The \emph{code2seq} can generate method names for invisible methods. The main reason is that it is not just to select some tokens from the input tokens to form a new method name, but to extract the features of AST corresponding to the method body and establish the relationship between these features and the tokens of the method name. If the AST structure of the given method $m_1$ and $m_2$ in the training set is highly similar, then the \emph{code2seq} probably recommends the method name token of $m_2$ to $m_1$, even if the token in the corresponding method name does not appear in the AST path token of $m_1$.

\begin{figure}[h]
\centering
\includegraphics[width=7.5cm]{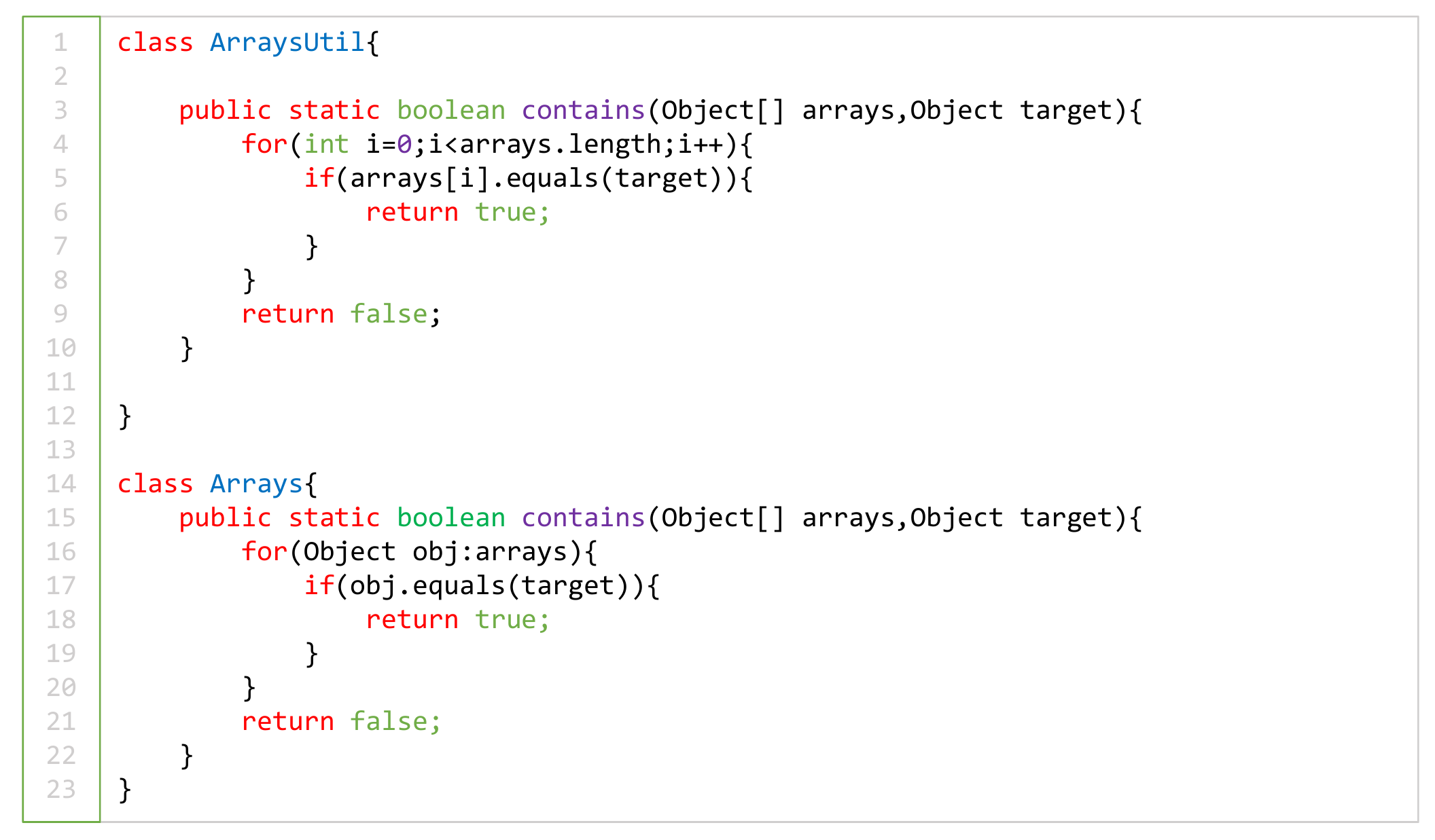}
\caption{Methods with similar AST trees.}
\label{fig:p1}
\end{figure}

A typical case is shown in the figure \ref{fig:p1}, where the \emph{code2seq} successfully recommends method names for invisible methods in the test set. We found that both visible and invisible methods have very similar AST trees. Moreover, the \emph{code2seq} can strongly identify the relationship between AST features and method name tokens, and favourably recommend the corresponding method names.

After the analysis in the previous figure, we can conclude that the performance of \emph{code2seq} in invisible methods slightly decreases. Still, it can accurately recommend method names.

\noindent \textbf{RQ3: Can the length of the method names recommended fulfill the standard of method name length?}

A method name is usually regarded as a brief description of the function of a piece of code. First, we expect that it can describe the function of a method in sufficient detail. Second, we hope it can help developers understand the function of a program quickly and concisely. Therefore, the length of the sequence contained in the method name recommended by the \emph{code2seq} should be within a reasonable range. Empirical studies show that the length of method name sequence in the range of 3 to 7 is the most conducive to improve the readability of the code, meanwhile the range conforms to the habit of human short-term memory.

To evaluate whether the length of the token sequence contained in the method name generated by \emph{code2seq} meets the custom of the method name length, we train \emph{code2seq} model on the training set derived from three datasets, \ie\ Liu-kui dataset, new-100 dataset, and new-400 dataset, respectively. Then, we test the method names generated on their corresponding test sets, and count the length distribution of the corresponding method name token sequences. 

%

As shown in Table \ref{tab:3-1}, the satisfaction rates are generated by the \emph{code2seq} on the test sets from three datasets (Liu-kui dataset, new-100 dataset, new-400 dataset).

\begin{table}[h]
\begin{tabular}{|c|c|c|c|}
\hline
                & \textbf{new-400 dataset} &\textbf{ Liu-kui dataset} & \textbf{new-100 dataset} \\ \hline
\textbf{Satisfaction}         & 216467          & 88889           & 2116           \\ \hline
\textbf{Dissatisfaction}      & 264572          & 122752          & 2594          \\ \hline
\textbf{Satisfaction rate}    & 0.44            & 0.41            & 0.44            \\ \hline
\end{tabular}
\caption{The Satisfaction rates on the three datasets.}
\label{tab:3-1}
\end{table}

\begin{figure}[h]
\centering
\includegraphics[width=8.5cm]{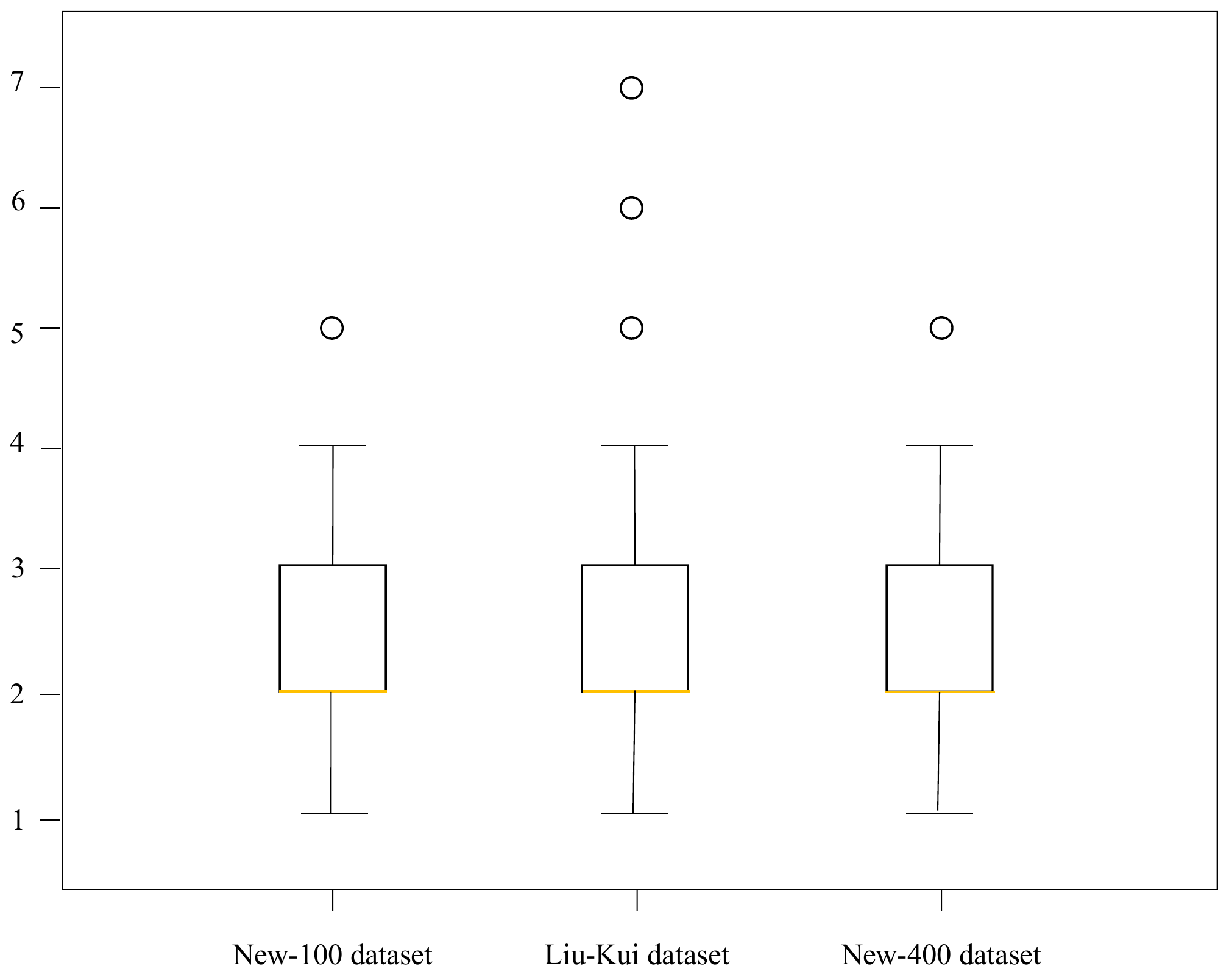}
\caption{The length distribution of code2seq's method name sequence generated on the three datasets.}
\label{fig:1}
\end{figure}

The box diagram of the length distribution of \emph{code2seq}'s method name sequence generated on the test set from three datasets (Liu-kui dataset, new-100 dataset and new-400 dataset) is shown in figure \ref{fig:1}. The length of most method name sequences generated by \emph{code2seq} on the three datasets is between 1 to 5, 1 to 7, and 1 to 5, respectively. According to the experimental results above, we can conclude that most of the method name sequences generated by \emph{code2seq} fulfill the standard of method name length.

We define satisfaction rate as follows.

\begin{equation}
\footnotesize
Satisfaction\ rate =\frac{N_{satisfaction}}{N_{satisfaction}+N_{dissatisfaction}}
\label{formu1}
\end{equation}

$N_{satisfaction}$ represents the number of methods that fulfill the standard, and $N_{dissatisfaction}$ means the number of methods that do not fulfill the standard. Satisfaction rate indicates the proportion of the number of methods that fulfill the standard of method name length to the total number of methods. The second to fourth columns show the number of methods that fulfill the standard of method name length ($N_{satisfaction}$), the number of methods that do not fulfill the standard of method name length ($N_{dissatisfaction}$), and the proportion of the number of methods that meet the length standard to the total number of methods (\emph{Satisfaction rate}).

From the table, we have the following findings:

\begin{itemize}
  \item Firstly, from the point of length distribution, the length of the method name sequences generated by \emph{code2seq} on the three datasets is in the range of 1 to 5, 1 to 7 and 1 to 5, respectively. It is true that the sequence length of most method names ranges from 1 to 4 tested on the three datasets. The result fulfills the standard of method name length.
  \item Secondly, in terms of \emph{satisfaction rate}, the number of method name sequences generated by \emph{code2seq} on the three datasets accounts for 44\%, 41\% and 44\%, respectively. Among the method names that do not fulfill the standard of method name length, the proportion of method names, whose length equals 2, is 32\%.
\end{itemize}

After the previous analysis, we can draw conclusion that most method names generated by the \emph{code2seq} fulfill the standard of method name length.

\section{Discussion of Related Work}
\label{sec4}

To study the performance of naming methods, we investigate the primary naming methods for the program code entity. One is to use machine translation methods such as neural-translation framework \emph{seq2seq} \cite{g3}, \emph{code2seq} \cite{g11}, and the method name is finally generated by translating the method body. The \emph{code2seq} \cite{g11} models a code snippet \cite{g25} as the set of compositional abstract syntax tree (AST) \cite{g10} paths and uses attention mechanism to select the relevant paths while decoding. Nguyen et al \cite{g3} presented \emph{MNire}, a machine learning approach to suggest a method name and to detect method name inconsistencies. First, they found high proportions of the tokens of method names can be found in the three contexts of a given method including its body, the interface (the method's parameter types and return type), and the enclosing class's name. Second, to suggest a good name for a method, they used \emph{seq2seq} to the naturalness of the program entities in the contexts.

Alon \cite{g10} presented a general path-based representation for predicting method names by combining abstract syntax tree (AST) \cite{g10}. Furthermore, Alon et al. proposed \emph{code2vec} \cite{g13} to represent a method body into code embeddings, \ie\ coverting method body into fixed-length, continuously distributed vectors, by aggregating the set of its corresponding AST paths with attentional network. The vectors of methods with similar AST structures are close to each other in the same space measured by semantic similarity among method names. The \emph{code2vec} can retrieve similar method bodies and reuse their method names. Jiang et al. \cite{g1} conducted an empirical study on the method name recommendation approach \emph{code2vec}. Their evaluation results suggest that \emph{code2vec} deserves significant improvement. However, to investigate the possibility of designing simple and straight-forward alternative approaches, they also proposed a heuristics based approach to recommending method names according to given method bodies. The experimental results show that their approach outperforms \emph{code2vec} significantly.

\section{Conclusions}
\label{sec5}

In this paper, we conduct an empirical study on such top-level method name recommendation approaches. Our evaluation suggests that we should adopt different naming approaches for different categories of method names. To intuitively reveal the advantages of such different naming approaches, we also empirically evaluate the \emph{code2seq}, \emph{HeMa}, and \emph{code2vec} especially in the part of naming categories, precision and recall.

\bibliographystyle{IEEEtran}
\bibliography{Reference}

\begin{thebibliography}{10}
\providecommand{\url}[1]{#1}
\csname url@samestyle\endcsname
\providecommand{\newblock}{\relax}
\providecommand{\bibinfo}[2]{#2}
\providecommand{\BIBentrySTDinterwordspacing}{\spaceskip=0pt\relax}
\providecommand{\BIBentryALTinterwordstretchfactor}{4}
\providecommand{\BIBentryALTinterwordspacing}{\spaceskip=\fontdimen2\font plus
\BIBentryALTinterwordstretchfactor\fontdimen3\font minus
  \fontdimen4\font\relax}
\providecommand{\BIBforeignlanguage}[2]{{%
\expandafter\ifx\csname l@#1\endcsname\relax
\typeout{** WARNING: IEEEtran.bst: No hyphenation pattern has been}%
\typeout{** loaded for the language `#1'. Using the pattern for}%
\typeout{** the default language instead.}%
\else
\language=\csname l@#1\endcsname
\fi
#2}}
\providecommand{\BIBdecl}{\relax}
\BIBdecl

\bibitem{g28}
R.~Oliveira, ``When more heads are better than one? understanding and improving
  collaborative identification of code smells,'' in \emph{IEEE/ACM 38th
  International Conference on Software Engineering Companion (ICSE-C)}, 2016,
  pp. 879--882.

\bibitem{g29}
J.~Czerwonka, M.~Greiler, and J.~Tilford, ``Code reviews do not find bugs. how
  the current code review best practice slows us down,'' in \emph{IEEE/ACM 37th
  IEEE International Conference on Software Engineering}, vol.~2, 2015, pp.
  27--28.

\bibitem{g27}
M.~Tufano, C.~Watson, G.~Bavota, M.~Di~Penta, M.~White, and D.~Poshyvanyk,
  ``Recovering clear, natural identifiers from obfuscated js names,'' in
  \emph{proceeding of the 11th Joint Meeting on Foundations of Software
  Engineering (ESEC/FSE)}, 2017, pp. 542--553.

\bibitem{g6}
S.~McConnell, ``Code complete,'' \emph{Pearon Education}, 2004.

\bibitem{g6-1}
K.~Beck, ``Clean code: a handbook of agile software craftsmanship,''
  \emph{Pearon Education}, 2009.

\bibitem{g6-2}
B.~Sharif and J.~I. Maletic, ``An eye tracking study on camelcase and
  under\_score identifier styles,'' in \emph{IEEE 18th International Conference
  on Program Comprehension}, 2010, pp. 196--205.

\bibitem{g30}
D.~W. Binkley, M.~Davis, D.~J. Lawrie, and C.~Morrell, ``To camelcase or
  under\_score,'' in \emph{IEEE/ACM 27th International Conference on Program
  Comprehension (ICPC)}, 2019, pp. 177--177.

\bibitem{g11}
\BIBentryALTinterwordspacing
U.~Alen, S.~Brody, O.~Levy, and E.~Yahav, ``Code2seq: Generating sequences from
  structured representations of code,'' in \emph{International Conference on
  Learning Representations of Code (ICLR)}, 2019, pp. 111--124. [Online].
  Available: \url{https://openreview.net/pdf?id=H1gKYo09tX}
\BIBentrySTDinterwordspacing

\bibitem{g3}
S.~Nguyen, H.~Phan, T.~Le, and T.~N. Nguyen, ``Suggesting natural method names
  to check name consistencies,'' in \emph{2020 IEEE/ACM 42nd International
  Conference on Software Engineering (ICSE)}, 2020, pp. 1372--1384.

\bibitem{g1}
L.~Jiang, H.~Liu, and H.~Jiang, ``Machine learning based recommendation of
  method names: How far are we,'' in \emph{2019 34th IEEE/ACM International
  Conference on Automated Software Engineering (ASE)}, 2019, pp. 602--614.

\bibitem{g13}
U.~Alon, M.~Zilberstein, O.~Levy, and E.~Yahav, ``code2vec: Learning
  distributed representations of code,'' in \emph{Proceedings of the ACM on
  Programming Languages}, 2019, pp. 1--27.

\bibitem{g31}
\BIBentryALTinterwordspacing
R.~S. Alsuhaibani, C.~D. Newman, M.~J. Decker, M.~L. Collard, and J.~I.
  Maletic, ``On the naming of methods: {A} survey of professional developers,''
  \emph{CoRR}, vol. abs/2102.13555, 2021. [Online]. Available:
  \url{https://arxiv.org/abs/2102.13555}
\BIBentrySTDinterwordspacing

\bibitem{g2}
K.~Liu, D.~Kim, T.~F. Bissyand, T.~Kim, K.~Kim, A.~Koyuncu, S.~Kim, and
  Y.~Le~Traon, ``Learning to spot and refactor inconsistent method names,'' in
  \emph{2019 IEEE/ACM 41st International Conference on Software Engineering
  (ICSE)}, 2019, pp. 1--12.

\bibitem{g25}
M.~Raghothaman, Y.~Wei, and Y.~Hamadi, ``Swim: Synthesizing what i mean: Code
  search and idiomatic snippet synthesis,'' in \emph{IEEE/ACM 38th
  International Conference on Software Engineering (ICSE)}, 2016, pp. 357--367.

\bibitem{g10}
U.~Alon, M.~Zilberstein, O.~Levy, and E.~Yahav, ``A general path-based
  representation for predicting program properties,'' in \emph{ACM 39th SIGPLAN
  Conference on Programming Language Design and Implementation}, 2018, pp.
  404--419.

\end{thebibliography}


\end{document}